\documentclass[10pt]{article}
\usepackage{geometry} 
\usepackage{graphicx}
\usepackage{amssymb,color}
\usepackage{cancel}

\def\eqref#1{(\ref{#1})}

\makeatletter
\@addtoreset{equation}{section}
\makeatother

\usepackage[colorlinks=true, pdfstartview=FitV, linkcolor=blue, citecolor=blue,urlcolor=blue]{hyperref}

\definecolor{darkgreen}{rgb}{0, 0.5, 0}
\definecolor{brown}{rgb}{0.4,0, 0}

\def\bea{\begin{eqnarray}}
\def\eea{\end{eqnarray}}
\def\le{\left}
\def \pa{\partial}
\def\ri{\right}

\newcommand{\eqa}{\begin{eqnarray}}
\newcommand{\eeqa}{\end{eqnarray}}
\newcommand{\beq}{\begin{equation}}
\newcommand{\eeq}{\end{equation}}

\newtheorem{dfn}{Definition}[section]
\newtheorem{thm}[dfn]{Theorem}
\newtheorem{rmk}[dfn]{Remark}
\newtheorem{lem}[dfn]{Lemma}

\newtheorem{emp}[dfn]{Example}

\newtheorem{prp}[dfn]{Proposition}

\newcommand{\nn}{\nonumber}

\newcommand{\p}{\partial}

\def\DF{!_{(r)}! }

\newcommand{\F}{\mathcal{F}}

\newenvironment{prf}{\noindent {\it Proof} \ }{\hfill $\Box$}

\title{The partition function of the extended $r$-reduced Kadomtsev-Petviashvili hierarchy}
\author{
{Marco Bertola$^{*,**,\dag}$, Di Yang$^{\dag}$}\\
{\small ${}^*$ Centre de recherches math\'ematiques, Universit\'e de Montr\'eal}\\
{\small C. P. 6128, succ. centre ville, Montr\'eal, Qu\'ebec, Canada H3C 3J7}\\
{\small ${}^{**}$ Department of Mathematics and Statistics, Concordia University,}\\
{\small 1455 de Maisonneuve W., Montr\'eal, Qu\'ebec, Canada H3G 1M8}\\
{\small ${}^{\dag}$ SISSA, via Bonomea 265, Trieste 34136, Italy}
}
\date{}
\begin{document}
\maketitle

\begin{abstract}
We derive a particular solution of the extended $r$-reduced KP hierarchy, which is specified by a generalized string equation. The work is a generalization to arbitrary $r\geq 2$ of Buryak's recent results of a solution to the extended open KdV hierarchy which corresponds to $r=2$.
\end{abstract}



\section{Introduction and results} 
Recently Pandharipande, Solomon and Tessler conjectured an integrable equation for descendent integrals for open and closed Riemann surfaces, called the open KdV equation \cite{PST} which has attracted significant attention \cite{Buryak1,Buryak2,Ke,A}. In particular, Buryak \cite{Buryak1,Buryak2} constructed an extension of this equation to a hierarchy of commuting flows, called the extended open KdV hierarchy; he also derived an explicit formula for a particular solution of the extended open KdV hierarchy which satisfies a generalized version of the string equation.

It is known that the partition function for $r$-spin structures of type $A$ is a particular tau function of the $r$-reduced KP hierarchy \cite{W,FSZ,FJR}.  
As far as we know, the open version of these invariants (except $r=2$, the open KdV case) has not been discovered or defined;  however, we are going to construct a particular solution of the so-called extended $r$-reduced KP hierarchy, for which we hope it will play the role of the partition function for open $r$-spin structures of type $A$.
In passing, we also present some formul\ae\ which clarify certain series expansions of \cite{Buryak2}.
To present the result, consider the differential operator $L$ 
\bea
 L=\p^{r}+\sum_{\alpha=1}^{r-1}u_\alpha(x)\,\p^{r-1-\alpha}\ ,\ \ \ \p:=\p_x, \ \ r\geq 2.
 \eea
 The $(r-1)$-KdV hierarchy, also called the Gelfand-Dickey hierarchy or the $r$-reduced KP hierarchy, is a hierarchy of evolutionary partial differential equations for the operator $L$ in terms of infinitely many auxiliary parameter $t_m$ (called customarily ``times'')
\begin{equation}\label{GD}
\frac{\p L}{\p t_m}=\alpha_m\,\left[\left(L^{\frac{m}{r}}\right)_+,L\right],\quad m\geq 1.
\end{equation}
This infinite system of evolution equations for $L$ (namely the coefficients $u_\alpha(x; t_1,\dots)$) is well--known to be {\em compatible} \cite{Dickii}. 
Here the numbers $\alpha_m$ are defined by 
\bea
\label{12}
\alpha_{p+q r}=\frac{1}{(p+qr)\DF},\quad 1\leq p\leq r,\,q\geq 0
\eea
and $x$ and $(p+qr)\DF $ is the $r_{th}$ generalized double factorial defined by
\bea 
(p+qr)\DF :=(p+qr)\cdot (p+(q-1)r)\cdot...\cdot p
\eea
 which reduces to the usual double factorial in the case $r=2$; the following convention is used
\bea
 0\DF= (-1)\DF =...=(-(r-1))\DF :=1.
 \eea
 The $m=1$ equation in \eqref{GD} reads $\frac {\p L}{\p t_1} = [\p, L]$, which means that $t_1$ simply acts by translations in the variable $x$; therefore we can (and will) identify $t_1$ and $x$ throughout the paper.
  Introduce the notations:
\bea
t=(t_1,...,t_{r-1},t_{r+1},...,t_{2r-1},...),\quad t_{_{KP}} =(t_1,t_2,t_3,...).
\eea
In other words, the infinite vector $t$ is obtained from $t_{_{KP}} $ by removing the variables with index which is a multiple of $r$. Note that if $m$ is a multiple of $r$, then the r.h.s. of \eqref{GD} vanishes\footnote{Indeed in this case the r.h.s is the commutator of $L$ with an {\em integer} power of $L$ itself.}  which implies that the operator $L$ and hence the functions $u_\alpha(t)$ depend only on the reduced set of variables $t$ (as opposed to the {\em extended} set $t_{_{KP}} $):
\bea
u_\alpha=u_\alpha(t),\quad \alpha=1,...,r-1.
\eea
Consider now the flows of the usual wave vector of the $r$--reduced KP hierarchy \eqref{GD}, but allow them to depend also  on the ``trivial'' flow variables $t_{rp}, p\in \mathbb N$:
\begin{equation}\label{ET}
\frac{\p \Psi}{\p t_m}=\alpha_m\left(L^{\frac{m}{r}}\right)_+\Psi,\quad m\geq 1.
\end{equation}
The compatibility of the equations \eqref{GD} , \eqref{ET} is a standard result (see for example \cite{Dickii}):
\bea
\p_{t_m}\p_{t_n}L=\p_{t_n}\p_{t_m}L,\quad
\qquad 
\p_{t_m}\p_{t_n}\Psi=\p_{t_n}\p_{t_m}\Psi,\quad  \forall\, m,n\geq 1.
\eea
So equations \eqref{GD} and equations \eqref{ET} together form an integrable system.
\begin{dfn}
\label{def11}
The infinite system of compatible equations \eqref{GD}, \eqref{ET}
 is called  the {\em extended $r$-reduced} KP hierarchy. 
 \end{dfn}
The case $r=2,$ coincides with the extended open KdV 
hierarchy \cite{Buryak1,Buryak2,PST}. We point out that we are not imposing at this stage an  additional spectral equation.

In this paper, we are going to introduce (in Theorem \ref{th1.1} below) a particular solution of the extended $r$-reduced KP hierarchy of Def. \ref{def11} that satisfies a certain additional constraint in the form of the {\bf string equation}. It turns out that this solution is closely related to a pair of adjoint ODE problems that are closely related to the Pearcey equation (see \eqref{o1}, \eqref{o2}); the approach also should shed light on certain objects that have appeared in the case $r=2$ in \cite{Buryak1}. 

\paragraph{Frobenius manifolds.} Recall now that the Frobenius manifold corresponding to the $r$-reduced KP hierarchy is the space of miniversal deformations of a simple singularity of type $A_{r-1}.$ Let $Z(t)$ denote the partition function \cite{AM,DZ-norm} of this Frobenius manifold. It is a particular tau function of the $r$-reduced KP hierarchy \eqref{GD} which is uniquely determined (up to a multplicative constant) by the celebrated {\em string equation} \cite{AM}
\bea 
\label{Seq}
L_{-1}Z(t)=0,
\eea
where 
\begin{equation}
\label{L-1}
L_{-1}:=\sum_{p\geq {r+1},r\nmid p}t_p\frac{\p }{\p t_{p-r}}+\frac{1}{2}\sum_{k=1}^{r-1} t_kt_{r-k}-\frac{\p}{\p t_1}.
\end{equation}
It can also be uniquely determined by using \textit{only} the $\mathcal{W}_{A_{r-1}}$-constraints \cite{AM,BM,LYZ}. According to Witten's $r$-spin conjecture \cite{W}, which was proved by Faber, Shadrin and Zvonkine in \cite{FSZ} and in a more general setting by Fan, Jarvis and Ruan in \cite{FJR}, $\log Z$ is also the generating function of intersection numbers over certain moduli space of $r$-spin structures of type $A$.

Our goal is to produce a solution of the extended $r$--reduced KP hierarchy which generalizes the above (see Theorem \ref{th1.1}) and the result of \cite{Buryak2}. To this end and to introduce the necessary objects, let us denote by $\psi(t_{_{KP}} ;z)$ the following wave function associated to $Z(t)$
\bea
\psi(t_{_{KP}} ;z)=\frac{Z(t-[z^{-1}])}{Z(t)}\,e^{\xi(t_{_{KP}} ;\,z)}
\label{psispectral}
\eea
where we have used the shorthand notations
\bea
\label{shift}
 (t_{_{KP}}  - [z^{-1}])  &:=& \le( t_1 - \frac{m_1}{z} , t_2- \frac {m_{2}}{z^2}, \dots\ri),\\
\label{phase}
\xi(t_{_{KP}} ;\,z)&:=&\sum_{k\geq 1} \alpha_k t_k z^k,\\
m_{p+qr}&:=&(p+(q-1)r)\DF ,\quad 1\leq p\leq r-1,\,q\geq 0.
\eea
The unwieldy normalizations of the time flows are chosen to simplify the comparison with existing results later on and it is also dictated by the application to the underlying geometric invariants.
The wave function $\psi(t_{_{KP}};z)$ is the ordinary wave function {\em with the spectral parameter $z$}, namely it satisfies 
\bea
L \psi(t_{_{KP}} ;z) = z^r  \psi(t_{_{KP}} ;z).
\eea
While $L$ is {\em independent} of the flows $t_p, t_{2p}, t_{3p}, \dots$, the wave function $\psi$ {\em does depend} on them, albeit in a trivial exponential way as dictated by \eqref{psispectral}. In order to emphasize this additional dependence we have used the subscript $t_{_{KP}} $ to indicate the dependence on the complete set of KP flows.

Now we are ready to state the result of this paper.

\begin{thm}\label{th1.1}
Let $\omega = {\rm e}^{\frac {i\pi}{r+1}}$ and set
\bea\label{D(z)-solution}
D(z;\Gamma)= -\frac{i \omega^{\frac {r+ 1}2}  }{\sqrt{2\pi}} z^{\frac {r+1} 2}  \exp\le(-\frac{ z^{r+1}}{r+1} \ri)  \int_{\omega^{-1}\Gamma}\!\! w^{-1}\,\exp\left[-\frac{w^{r+1}}{(r+1)r} + \frac w r z^r\right] d w
\eea
where $\Gamma = {\rm e}^{-\frac{i \pi}{r+1}} \mathbb R_+ \cup {\rm e}^{\frac{i \pi}{r+1}} \mathbb R_+$ (traversed in the upward direction). It has the following asymptotic expansion
\bea
\label{Dasymp}
D(z;\Gamma)\sim 1+\sum_{k=1}^\infty \frac{d_k}{z^{(r+1)k}}=:d(z),\quad z\rightarrow \infty,\,z\in S'
\eea
where $S'=\{-\frac{\pi}{r}-\frac{\pi}{r+1}<arg(z)<\frac{\pi}{r}-\frac{\pi}{r+1}\}$ and $d_k$ are constants, and the contour has zero index number relative to $w=0$.
Introduce the following function 
\begin{equation}
\Psi(t_{_{KP}} ):=\frac{1}{2\pi i}\,\oint_{z=\infty} d(z)\, \psi(t_{_{KP}} ;z)\,\frac{dz}{z}
\end{equation}
where the  formal series $d(z)$ is given in \eqref{Dasymp}. Define also the {\em partition function} of the extended $r$--reduced KP hierarchy by the formula 
\begin{equation}
Z_E(t_{_{KP}} ):=Z(t)\,\Psi(t_{_{KP}} )
\end{equation}
where $Z(t)$ is the partition function of the $A_{r-1}$ Frobenius manifold satisfying the string equation \eqref{Seq}. Then we have
\bea \label{estr}
&&L_{-1}^{ext}\, Z_E(t_{_{KP}} )=0,\\
\label{Psi-ini} 
&&\Psi(t_{_{KP}} )|_{t_{KP\geq 2}=0}\equiv 1  \ \ (\forall\, t_1)
\eea
where the operator $L_{-1}^{ext}$ is defined by
\bea L_{-1}^{ext}=\sum_{p\geq {r+1}}t_p\frac{\p }{\p t_{p-r}}+\frac{1}{2}\sum_{k=1}^{r-1} t_kt_{r-k}+ t_{r}-\frac{\p}{\p t_1}. 
\eea

\noindent Furthermore, the solution $\Psi(t_{_{KP}} )$  of \eqref{ET} and such that  $Z_E(t_{_{KP}} ):=Z(t)\cdot \Psi(t_{_{KP}} )$ solves \eqref{estr} is unique up to a multiplicative constant.
\end{thm}

The paper is organized as follows: In Section 2, we review the tau function of the ($r$-reduced) KP hierarchy and the associated wave function and dual wave function.
In Section 3, by applying the string actions we study some properties of the wave function and the dual wave function associated to the partition function of the $r$-spin structures of type $A$. In Section 4, we prove Theorem \ref{th1.1}.
 
\section{Tau function and wave functions of the $r$-reduced KP hierarchy}
Let us consider solutions to the extended $r$-reduced KP hierarchy \eqref{GD},\eqref{ET}. It is easy to see that equations \eqref{GD} and \eqref{ET} are decoupled: we can  first solve \eqref{GD} and then solve \eqref{ET}. Equations \eqref{GD} have a class of well-known solutions corresponding to infinite Grassmannians. 

Let $u_1(t),...,u_{r-1}(t)$ be any Grassmannian solution to the $r$-reduced KP hierarchy. It is known that there exists a so-called Sato tau function $\tau(t_{_{KP}} )$ of this solution, which satisfies the bilinear identities
\bea
\frac{1}{2\pi i}\,\oint_{z=\infty} e^{\xi(t_{_{KP}} -t_{_{KP}} ';\,z)} \, \tau(t_{_{KP}} -[z^{-1}])\,\tau(t_{_{KP}} '+[z^{-1}])\, dz=0,\quad \forall \,t_{_{KP}} ,t_{_{KP}} '.
\label{SatoBil}
\eea
where $\xi(t_{_{KP}} -t_{_{KP}} ';z)$ and $t_{_{KP}} -[z^{-1}],\,t_{_{KP}} +[z^{-1}]$ are defined as in \eqref{phase},\,\eqref{shift}. The tau function $\tau$ is unique up to a factor of form
\begin{equation}
\label{factor}
\exp\left\{\sum_{k\geq 1} c_k t_k+c_0\right\}
\end{equation}
where $c_i,\,$ are arbitrary constants.

Substituting the solution $u_1(t),...,u_{r-1}(t)$ into equations \eqref{ET} we get a set of evolutionary linear PDEs for $\Psi.$ To solve them, let us introduce an auxiliary problem
\begin{equation}\label{spec}
L\psi( t_{_{KP}};z)=z^{r}\psi(t_{_{KP}};z).
\end{equation}
Together with the original PDEs \eqref{ET}
\bea
\frac{\p \psi}{\p t_m}(t_{_{KP}};z)=\alpha_m\left(L^{\frac{m}{r}}\right)_+\psi(z; t_{_{KP}}),\quad m\geq1
\eea 
$\psi( t_{_{KP}};z)$ becomes the wave function of the solution with the spectral parameter $z$ (also called the  Baker--Akhiezer function). Without loss of generality, we assume it is normalized in the following way
\bea\label{asym-psi}
\psi(t_{_{KP}} ;z)\sim \left(1+\mathcal{O}(z^{-1})\right)e^{\xi(t_{_{KP}} ;z)},\qquad z\rightarrow \infty.
\eea
Note that the wave function is unique up to an arbitrary asymptotic series $g(z)$ satisfying
\begin{equation}\label{gz}
g(z)\sim1+\sum_{k=1}^\infty \frac{g_k}{z^k},\quad z\rightarrow \infty
\end{equation}
where $g_k,\,k\geq 1$ are constants. The following lemma is well-known.

\begin{lem} (\cite{DJMK,Dickii})
For any Grassmannian solution $u_1(t),...,u_{r-1}(t)$ of the $r$-reduced KP hierarchy,  let $\tau(t_{_{KP}} )$ be any tau-function of this solution. Then
\bea
\psi(t_{_{KP}} ;z)=\frac{\tau(t_{_{KP}} -[z^{-1}])}{\tau(t_{_{KP}} )}e^{\xi(t_{_{KP}} ;\,z)}.
\eea
is a wave function of the solution.
\end{lem}

Similarly to the above construction, let us also recall the \textit{dual} wave function $\psi^\star(t_{_{KP}} ;z)$ of a solution $u(t).$ It satisfies the following linear system:
\bea
&&L^\star \psi^*=z^r \psi^*,\\
&&\frac{\p\psi^*}{\p t_m}=-\alpha_m \left(L^{m/r}\right)_+^\star\, \psi^*.
\eea
Here $\star$ denotes the formal adjoint operator. The dual wave function is also normalized such that
\bea\label{asym-psi-star}
\psi^*(t_{_{KP}} ;z)=\left(1+\mathcal{O}(z^{-1})\right)e^{-\xi(t_{_{KP}} ;z)},\qquad z\rightarrow \infty.
\eea
In terms of the tau function it is expressed by
\bea
\psi^*(t_{_{KP}} ;z):=\frac{\tau(t_{_{KP}} +[z^{-1}])}{\tau(t_{_{KP}} )}e^{-\xi(t_{_{KP}} ;\,z)}.
\eea
The bilinear identities \eqref{SatoBil} read as follows
\bea\label{bilinear-psipsi'}
\frac{1}{2\pi i}\,\oint_{z=\infty} \psi(t_{_{KP}} ;z)\,\psi^*(t_{_{KP}}' ;z)\, dz=0,\quad \forall \,t_{_{KP}} ,t_{_{KP}} '.
\eea

Now we take a tau function $\tau(t)$ of the solution $u_1(t),...,u_{t-1}(t)$ without dependence in $t_{rp},\,p\geq 1,$ and take $g(z)$ any asymptotic series of form \eqref{gz}. Then solutions of equations \eqref{ET} are given by
\begin{equation}\label{PSI}
\Psi(t_{_{KP}} )=\frac{1}{2\pi i}\,\oint_{z=\infty} g(z)\, \frac{\tau(t-[z^{-1}])}{\tau(t)}e^{\xi(t_{_{KP}} ;\,z)}\,\frac{dz}{z}.
\end{equation}
Inspired by Buryak \cite{Buryak2} we call 
\bea 
\tau_E(t_{_{KP}} ):=\tau(t)\,\Psi(t_{_{KP}} )
\eea a tau function of the solution 
$(u_1(t),...,u_{r-1}(t),\Psi(t_{_{KP}} ))$ of the extended $r$-reduced KP hierarchy.

It is well-known that the $r$-reduced KP hierarchy  is Miura equivalent to Dubrovin-Zhang's integrable hierarchy of topological type associated to the $A_{r-1}$ Frobenius manifold. So we can write the hierarchy \eqref{GD} by using the \textit{normal coordinates} $w_1,...,w_{r-1}$ \cite{DZ-norm, DZ1}
\bea
\frac{\p w_\alpha}{\p t^{\beta,q}}=\eta_{\alpha\gamma}\p_x\frac{\delta \bar{h}_{\beta,q}}{\delta w_\gamma(x)},\quad q\geq 0
\eea
where $\eta_{\alpha\beta}=\delta_{\alpha+\beta,\,r},$ $\frac{\delta}{\delta w_\gamma}$ denote the variational derivatives, $\bar{h}_{\beta,q}=\int h_{\beta,q}\,dx$ are Hamiltonians whose densities $h_{\beta,q}$ satisfy the tau-symmetry property
\bea
h_{\alpha,-1}=w_\alpha,\quad \frac{\p h_{\beta,p-1}}{\p t^{\alpha,q}}=\frac{\p h_{\beta,q-1}}{\p t^{\alpha,p}},\quad p,q\geq 0.
\eea
In these notations, $t^{\alpha,p}=t_{\alpha+p r},\, p\geq 0.$ Here and below free Greek indices always take values $1,...,r-1,$ the Einstein summation convention for repeated Greek indices with one-up and one-down is always assumed, and we use $\eta_{\alpha\beta}$ and its inverse $\eta^{\alpha\beta}$ to lower and raise Greek indices. For example, $w^\alpha:=\eta^{\alpha\beta}w_\beta.$

The tau symmetry property allows us to define two-point functions \cite{DZ-norm} $\Omega_{\alpha,p;\beta,q}$ for an arbitrary solution $(w_1(t),...,w_{r-1}(t))$
\bea
\Omega_{\alpha,p;\beta,q}(w;w_x,w_{xx},...)=\p_x^{-1}\left(\frac{\p h_{\beta,p-1}}{\p t^{\alpha,q}}\right),\quad p,q\geq 0.
\eea
It also implies the existence of a tau function $\tau$ of the solution such that
\bea\frac{\p^2\log \tau}{\p t^{\alpha,p} \p t^{\beta,q}}=\Omega_{\alpha,p;\beta,q}(w(t);w_x(t),w_{xx}(t),...).\eea
We call this tau function the {\em Dubrovin-Zhang tau function} associated to the $A_{r-1}$ Frobenius manifold.
Note that $\Omega_{\alpha,p;\beta,q}$ are differential polynomials in $w$, and that $\log \tau$ admits the genus expansion
\bea\log \tau=\sum_{g=0}^\infty \F_g=:\F,\eea
where $\F_g$ are genus $g$ free energies and $\F$ is the free energy. In the particular example of the $r$-reduced KP hierarchy \eqref{GD} that we are considering, the normal coordinates $w_\alpha$ as well as the Hamiltonian densities $h_{\alpha,p}$ are given by \cite{DJMK, DZ-norm}
\bea
\label{normal-lax} w_\alpha&=&\frac{h_\alpha }{\alpha\DF },\\
\label{density-lax} h_{\alpha,\,p}&=&\frac{h_{\alpha+(p+1)r}}{(\alpha+(p+1)r)\DF }
\eea
where
\bea \label{h_k-lax} h_k:=Res\,L^{\frac{k}{r}},\qquad k\geq 1. \eea
In the above formula, $Res$ means taking the coefficient of $\p^{-1}.$

For Grassmannian solutions, the Sato tau function and the Dubrovin-Zhang tau function coincide \cite{DJMK,DZ-norm} modulo an arbitrary factor of form \eqref{factor}.
\section{Wave potentials, string actions and Pearcey integrals}
Recall that the partition function $Z(t)$ associated to the $A_{r-1}$ Frobenius manifold is a particular tau function of the $r$-reduced KP hierarchy. Up to a multiplicative constant, this tau function is uniquely specified by the string equation
\begin{equation}\label{str}
L_{-1}\,Z=0
\end{equation}
where $L_{-1}$ is defined in \eqref{L-1}.
Write 
\bea \langle\tau_{k_1}...\tau_{k_m}\rangle:=\frac{\p^m\log Z}{\p t_{k_1}...\p t_{k_m}}\Big|_{t=0}. \eea
Then according to \cite{FSZ,LYZ}, we know that $\langle\tau_{k_1}...\tau_{k_m}\rangle$ vanishes unless there exists a non-negative number $g$ such that
\bea \frac{k_1}{r}+...+\frac{k_m}r=m+\left(\frac{r-2}{r}-3\right)(1-g). \eea
This implies for $r\geq 3,$
$Z(x,0,0,...)=constant.$ As usual we normalize this constant to be 1. For $r=2$, it is known that $Z(x,0,0,...)=\exp\left(x^3/6\right).$

We denote by $u_1(t),...,u_{r-1}(t)$ the solution to the $r$-reduced KP hierarchy associated with the tau function $Z(t),$ in normal coordinates by $w_1(t),...,w_{r-1}(t).$ The string equation \eqref{str} implies of the initial data
\bea w_\alpha|_{t_{\geq 2}=0}=\frac{\p^2\log Z}{\p t_1 \p t_\alpha}\Big{|}_{t_{\geq 2}=0}=\delta_{\alpha,\,r-1}\cdot x,\quad \alpha=1,...,r-1.\eea
Here $\delta_{\alpha,\,r-1}$ is the Kronecker delta function.
\begin{lem}
The Miura transformation relating $u$ to $w$ is triangular and of form
\bea w_\alpha=\frac{1}{r} \, u_\alpha+M_\alpha(u_1,...,u_{\alpha-1}),\eea
where $M_\alpha$ are differential polynomials in $u_1,...,u_{\alpha-1}.$
\end{lem}
\begin{prf}
Let us introduce  a grading on pseudo-differential operators:
\bea \deg u_{\alpha}^{(k)}:=\alpha+1+k,\quad \deg \p^m:=m,\qquad k\geq 0,\,m\in\mathbb{Z}.\eea
Then we find that $L$ and $L^{\frac1r}$ are homogeneous operators with degree
\bea \deg L=r,\quad \deg L^{\frac1r}=1.\eea
Recall due to \eqref{normal-lax},\,\eqref{h_k-lax} that 
\bea w_\alpha=\frac{1}{\alpha\DF } Res L^{\frac{\alpha}r}.\eea
So we have
\bea \deg w_\alpha=\alpha+1.\eea
The leading term $\frac{1}{r}u_\alpha$ is an easy exercise: one can replace in $L$ the operator $\p$ with a parameter $\lambda$ and take the usual residue at $\lambda=\infty.$  The lemma is proved.
\end{prf}

The above Lemma implies
\begin{equation}
u_\alpha|_{t_{\geq 2}=0}=\frac{1}{r}\,\delta_{\alpha,r-1}\cdot x,\quad \alpha=1,...,r-1.\label{initial}
\end{equation}
In particular, we have
\begin{lem}\label{iiinitial-u}
The following initial values hold
\bea
 u_\alpha|_{t=0}=0. \label{all-t-ini} 
\eea
\end{lem}

Denote by $\psi(t_{_{KP}} ;z)$ and $\psi^*(t_{_{KP}} ;z)$ the wave and the dual wave functions associated to $Z(t)$, respectively:
\bea
\psi(t_{_{KP}} ;z)=\frac{Z(t-[z^{-1}])}{Z(t)}e^{\xi(t_{_{KP}} ;\,z)},\quad \psi^*(t_{_{KP}} ;z)=\frac{Z(t+[z^{-1}])}{Z(t)}e^{-\xi(t_{_{KP}} ;\,z)},
\eea
and introduce the following notations
\bea
K(t_{_{KP}} ;z)&:=&Z(t)\cdot \psi(t_{_{KP}} ; z),\qquad K^*(t_{_{KP}} ;z):=Z(t)\cdot \psi^*(t_{_{KP}} ; z),\\
f(x;z)&:=&\psi(t_{_{KP}} ;z)|_{t_{\geq 2}=0},\qquad f^*(x;z):=\psi^*(t_{_{KP}} ;z)|_{t_{\geq 2}=0}.\label{fandfstar}
\eea
We call $K(t_{_{KP}} ;z)$ and $K^*(t_{_{KP}} ;z)$ the \textit{wave potential} and the \textit{dual wave potential} associated to the partition function $Z(t).$
Let us consider the string actions on them.
\begin{prp} \label{string-action-wave} The following equalities hold true:
\bea
L_{-1}^{ext}K(t_{_{KP}} ;z)&=&S_z K(t_{_{KP}} ;z),\label{grass}\\
L_{-1}^{ext,*}K^*(t_{_{KP}} ;z)&=&-S_z^\star K^*(t_{_{KP}} ;z).\label{grass-pair}
\eea
Here $S_z$ is defined by
\bea\label{def-Sz}
S_z=z^{-(r-1)/2}\circ \p_z\circ z^{-(r-1)/2}-z=\frac{1}{z^{r-1}}\p_z-\frac{r-1}{2\,z^r}-z
\eea
and $S_z^\star$ is the formal adjoint operator of $S_z$ which has the form
\bea\label{def-Sz-star}
S^\star _z =z^{-(r-1)/2}\circ \p_z^\star \circ z^{-(r-1)/2}-z=-\frac{1}{z^{r-1}}\p_z+\frac{r-1}{2\,z^r}-z,
\eea
and $L_{-1}^{ext},\, L_{-1}^{ext,*}$ are defined by
\bea 
L_{-1}^{ext}&=&\sum_{p\geq {r+1}}t_p\frac{\p }{\p t_{p-r}}+\frac{1}{2}\sum_{k=1}^{r-1} t_kt_{r-k}+ t_{r}-\frac{\p}{\p t_1},\\
L_{-1}^{ext,*}&=&\sum_{p\geq {r+1}}t_p\frac{\p }{\p t_{p-r}}+\frac{1}{2}\sum_{k=1}^{r-1} t_kt_{r-k}-t_{r}-\frac{\p}{\p t_1}.
\eea
\end{prp}
\begin{prf}
On one hand, by using the string equation \eqref{str} we find
\bea
L_{-1}^{ext}K (t_{_{KP}} ;z)&=&\sum_{p\geq r+1}\frac{m_p}{z^p}\frac{\p Z(t-[z^{-1}])}{\p t_{p-r}}e^{\xi(t_{_{KP}} ;z)}-\frac{1}{2}\frac{r-1}{z^r} K(t_{_{KP}} ;z)-zK(t_{_{KP}} ;z)\nn\\
&&+\sum_{p\geq r+1}\alpha_{p-r}t_pz^{p-r} K(t_{_{KP}} ;z)+\sum_{p=1}^{r-1}m_{r-p}\,t_pz^{p-r}K(t_{_{KP}} ;z)+t_r K(t_{_{KP}} ;z).\nn\\
\eea
On another hand from the chain rule it follows that 
\bea 
\frac{1}{z^{r-1}}\p_z K(t_{_{KP}} ;z)=\sum_{p\geq r+1}(p-r)\frac{m_{p-r}}{z^p}\frac{\p Z(t-[z^{-1}])}{\p t_{p-r}}e^{\xi(t_{_{KP}} ;z)}+\sum_{p\geq 1} p\,\alpha_p\,t_pz^{p-r}K(t_{_{KP}} ;z). 
\eea
Comparing the above two equalities we have
\bea 
L_{-1}^{ext}K(t_{_{KP}} ;z) =\frac{1}{z^{r-1}}\p_z K(t_{_{KP}} ;z) -\frac{r-1}{2\,z^r}K(t_{_{KP}} ;z)-zK(t_{_{KP}} ;z)=S_z K(t_{_{KP}} ;z).
\eea
In a similar way we can prove \eqref{grass-pair}. The lemma is proved.
\end{prf}

\begin{lem}\label{f,f*} Let $f,f^*$ as in \eqref{fandfstar}. We have for any $k\geq 0,$
\bea
\p_x^k f(x;z)&=&(-1)^k S_z^k f(x;z),\label{expand-f}\\
\p_x^k f^*(x;z)&=&\left(S_z^\star\right)^k f^*(x;z).\label{expand-f*}
\eea
\end{lem}
\begin{prf}
Note that in the case $r=2$ we have $Z(x,0,0,...)=\exp\left(\frac{x^3}{6}\right)$ and that in the case $r\geq 3$ we have $Z(t_1,0,0,...)=1.$
Taking $t_2=t_3=....=0$ in both sides of \eqref{grass}, we find
\bea f_x+S_z f=0 \eea which implies \eqref{expand-f}. Similarly, taking $t_2=t_3=...=0$ in both sides of \eqref{grass-pair}, we obtain \eqref{expand-f*}.
\end{prf}

\begin{lem}\label{Taylor-ff*}
Denote $A(z)=f(0;z)$ and $A^*(z)=f^*(0;z)$. Applying the Taylor expansion and Lemma \ref{f,f*} we have
\bea 
\label{Taylor-A} 
f(x;z)&=&\sum_{n=0}^\infty \frac{(-1)^n}{n!} S_z^n A(z)\, x^n,\\
f^*(x;z)&=&\sum_{n=0}^\infty \frac{1}{n!} {S_z^\star}^n A^*(z)\, x^n.
\eea
\end{lem}

\begin{lem}\label{defining-A-A*} The functions $A(z),A^*(z)$ satisfy the following ODEs:
\bea 
S_z^r A(z)&=&(-z)^r A(z),\label{o1}\\
{S_z^\star}^r A^*(z)&=&(-z)^r A^*(z).\label{o2}
\eea
\end{lem}
\begin{prf}
Recall that the Lax equations for $\psi(t_{_{KP}} ;z)$ and $\psi^*(t_{_{KP}} ;z)$ read
\bea L\psi=z^r\psi,\qquad L^\star\psi^*=z^r \psi^*.\eea
Taking $t=0$ in these two equations and using the initial data \eqref{all-t-ini} we have
\bea \p_x^r f(x;z)|_{x=0}=z^r f(0;z),\qquad (-1)^r f^*(x;z)|_{x=0}=z^r f(0;z).\eea
Finally by employing Lemma \ref{f,f*} we obtain
\bea (-1)^r S_z^r f(x;z)|_{x=0}=z^r f(0;z),\qquad (-1)^r {S_z^*}^r f^*(x;z)|_{x=0}=z^r f^*(0;z).\eea
The lemma is proved.
\end{prf}

Note that \eqref{asym-psi},\,\eqref{asym-psi-star} require the (formal) boundary behaviour
\bea 
\label{Asym-1} A(z)&=& 1+\mathcal{O}(z^{-1}),\qquad z\rightarrow \infty,\\
\label{Asym-2} A^*(z)&=& 1+\mathcal{O}(z^{-1}),\qquad z\rightarrow \infty.
\eea
In order to solve the ODEs \eqref{o1},\,\eqref{o2} let us introduce a change of variable $\eta=z^r$
and let 
\bea 
H(\eta)=\eta^{\frac{r-1}{2r}}\exp\left(\frac{1}{r+1}\eta^{\frac{r+1}r}\right) = 
z^{\frac {r-1}2} \exp \le( \frac{ z^{r+1}}{r+1}\ri).
\eea
A straightforward computation shows that
\bea 
\label{zxi}
\frac1H\circ S_z\circ H=r\,\p_\eta,
\eea
which allows to transform immediately the ODEs \eqref{o1}, \eqref{o2} in a pair of adjoint Pearcey equations. Indeed, let us write $A(z)=H(\eta)\cdot \ell(\eta)$: the above ODE \eqref{o1} becomes
\bea 
(r\p_\eta)^r \ell(\eta)=(-1)^r\eta\, \ell(\eta).
\eea
Its solutions are well known and are given in the form of a Fourier-Laplace integral representation
\bea 
\ell(\eta)=C_1\, \int_\Gamma \exp\left(\frac{w^{r+1}}{(r+1)r}-\frac{w}{r}\eta\right) dw
\eea
where the contour $\Gamma$ is any contour extending to infinity along two different asymptotic directions in the sectors where $\Re (w^{r+1})$ tends to $-\infty$, and $C_1$ is a constant. Applying a formal saddle point method to meet the condition \eqref{Asym-1}, we obtain
\bea
\label{Pearcey}
A(z;\Gamma) = \frac{i}{\sqrt{2\pi}} z^{\frac {r-1} 2}  \exp\le( \frac{ z^{r+1}}{r+1} \ri)  \int_{\Gamma} \exp\left[\frac{w^{r+1}}{(r+1)r} - \frac w r z^r\right] \ d w
\eea
where the contour $\Gamma$ is any contour extending to infinity along two different asymptotic directions  $\mathcal R_j :=\mathbb R_+ {\rm e}^{\frac {i\pi}{r+1} + \frac {2i\pi}{r+1} j}$, $j\equiv 0,\dots, r\, {\rm mod}\, (r+1)$.  The integrals appearing in \eqref{Pearcey} (in the variable $\xi = z^r$) are famously known as {\em Pearcey} integrals and generalize the usual integral representation of Airy functions. A relatively straightforward steepest descent analysis shows that if $\Gamma$ is the contour originating at $\infty$ along $\mathcal R_{-1}$ and ending at $\infty$ along $\mathcal R_{0}$ then the function has the asymptotic expansion
\bea
A(z;\Gamma )=1+\mathcal{O}(1/z),\quad |z|\rightarrow \infty,\,z\in S_\Gamma,
\eea
where $S_\Gamma=\{|arg(z)|<\frac{\pi}{r}\}.$  More precisely the expansion contains only powers of $z^{-r-1}$ as stated in the following lemma (whose proof is straightforward and thus omitted).
\begin{lem} \label{A(z)-asymp}
$A(z;\Gamma)$ admits the following asymptotic
\bea A(z;\Gamma)\sim 1+\sum_{k=1}^\infty \frac{a_k}{z^{(r+1)k}},\qquad z\rightarrow \infty,\,z\in S_{\Gamma},\eea
where $a_k$ are constants.
\end{lem}
Below we denote by $a(z)=1+\sum_{k=1}^\infty \frac{a_k}{z^{(r+1)k}}$ the formal asymptotic series  of $A(z;\Gamma)$.
\begin{lem}\label{rotate}
With $\omega = {\rm e}^{\frac {i\pi}{r+1}}$, we have
\bea 
S_z^\star=\omega^{-1}S_{\omega z}.
\eea
\end{lem}
\begin{prf}
\bea S_{\omega z}=\frac{1}{\omega^r\,z^{r-1}}\p_z-\frac{r-1}{2\,\omega^r\, z^r}-\omega z=\omega\left(-\frac{1}{z^{r-1}}\p_z+\frac{r-1}{2\,z^r}-z\right)=\omega\,S_z^\star.\eea
\end{prf}

Lemma \eqref{rotate} implies
\bea 
A^*(z;\Gamma)&=&A(\omega z;\Gamma)\nn\\
&=& \frac{i \omega^{\frac {r+ 1}2}  }{\sqrt{2\pi}} z^{\frac {r-1} 2}  \exp\le( -  \frac{ z^{r+1}}{r+1} \ri)  \int_{\omega^{-1}\Gamma} \exp\left[\frac{ -s ^{r+1}}{(r+1)r} +  \frac s {r }  z^r\right] \ d s
\eea
which satisfies
\bea 
A^*(z;\Gamma)=1+\mathcal{O}(z^{-1}),\qquad z\rightarrow \infty,\,z\in \omega S_\Gamma.
\eea
We recall now that there exists a natural bilinear pairing between solutions of two adjoint equations which is called the {\bf bilinear concomitant}. In the present case, solutions of the two equations can be represented in terms of Fourier--Laplace integrals along two relatively dual bases in a certain homology of contours and the bilinear concomitant is simply a representation of the intersection pairing. For more details (and in more general terms) see  
 \cite{Bertola:semiclassBOPsJAT}, Sect. 3. Let us call 
\bea
B(z;\hat \Gamma) = C  z^{\frac {r-1} 2}  \exp\le( -  \frac{ z^{r+1}}{r+1} \ri)  \int_{\hat \Gamma} \exp\left[\frac{ - s^{r+1}}{(r+1)r} +  \frac s {r }  z^r\right] \ d s, 
\eea
where $\hat \Gamma$ spans the dual homology of the contours spanned by $\Gamma$'s.
The concomitant reads:
\bea
\label{BC}
\mathcal B( A(z;\Gamma), B(z;\hat \Gamma)) = 
\int_{\Gamma \times \hat \Gamma}  d s \, d w \frac {w^r-s^r}{w-s} \exp \le[
\frac {w^{r+1}- s^{r+1}}{r (r+1)}  - \frac {w -s}r \eta 
\ri].
\eea
Here $\eta = z^r.$
The result of this integral is the intersection number of $\Gamma$ and $\hat \Gamma$. Indeed
\bea
&&\int_{\Gamma \times \hat \Gamma}  d s \, d w \frac {w^r-s^r}{w-s} \exp \le[
\frac {w^{r+1}- s^{r+1}}{r (r+1)}  - \frac {w -s}r \xi 
\ri] \nn\\
&&= \int_{\Gamma \times \hat \Gamma}  d s \, d w \le(\pa_s + \pa _w\ri ) \frac{ \exp \le[
\frac {w^{r+1}- s^{r+1}}{r (r+1)}  - \frac {w -s}r \xi 
\ri]}{w-s}= 2i \pi \,link(\Gamma,\hat\Gamma).
\eea
The bilinear concomitant \eqref{BC} can equivalently be written in terms of a bilinear expression in the derivatives of the functions $A(z;\Gamma)$ and $B(z;\hat \Gamma)$ by expanding $\frac {w^r-s^r}{w-s} = \sum_{j=0}^{r-1} w^j s^{r-1-j}$ and thus giving the following lemma:
\begin{lem} \label{concomitant}
The following identities hold true:
\bea  \sum_{k=0}^{r-1} {S_z^\star}^{r-1-k} \left(A(\omega z;\hat \Gamma)\right) S_z^k A(z;\Gamma)=link(\Gamma,\hat\Gamma)\cdot z^{r-1},\quad z\in S_\Gamma.
\eea
See Sect. 3 of \cite{Bertola:semiclassBOPsJAT}.
\end{lem}
We remark that the analytic function $A(z;\Gamma)$ that we have defined has different asymptotic behaviours at $\infty$ as the sector changes. This is called the Stokes phenomenon.

To end this section, let us solve the following ODE problem which will be useful for the next section:
\bea
 -S_z^\star\left(\frac{D(z)}{z}\right)=A(\omega z)
 \label{DA}
\eea
with the (formal) asymptotic boundary  requirement
\bea 
\label{450}
D(z)=1+\mathcal{O}(z^{-1}),\qquad z\rightarrow\infty.
\eea

Using the representation \eqref{zxi} for the differential operator, it is immediate to verify that  
\bea
\label{Dzsol}
D(z;\Gamma)= -\frac{i \omega^{\frac {r+ 1}2}  }{\sqrt{2\pi}} z^{\frac {r+1} 2}  \exp\le(-\frac{ z^{r+1}}{r+1} \ri)  \int_{\omega^{-1}\Gamma}\!\! w^{-1}\,\exp\left[-\frac{w^{r+1}}{(r+1)r} + \frac w r z^r\right] d w.
\eea
(which is \eqref{D(z)-solution}) indeed solves the equation. Of course the equation \eqref{DA} only determines the solution up to addition of the complementary solution of the associated homogeneous equation. This arbitrariness is reflected in the choice of the contour in \eqref{Dzsol}; indeed the integrand has a simple pole at $w=0$ and thus it is important to specify the index of the contour of integration relative to $w=0$.  Now, the steepest descent analysis shows that \eqref{Dzsol} has the behaviour \eqref{450} {\em provided} that the contour can be retracted onto the steepest descent path of the exponential phase. The steepest descent path clearly recedes to infinity (the principal saddle point is at $w = z$). Thus the contour must have index $0$ relative to $w=0$. 
\begin{rmk}
Another simple argument to determine the index of the integration contour relative to $w=0$ is that if the contour in \eqref{Dzsol} has index $1$, then  retracting the contour to the saddle yields the residue at $w=0$, which adds a multiple of ${\sqrt{2\pi}} z^{\frac {r+1} 2}  \exp\le(-\frac{ z^{r+1}}{r+1} \ri) $ (the complementary solution), which has necessarily an exponential growth within the sector $S_\Gamma$, and thus violates the boundary condition \eqref{450}.
\end{rmk}

 Applying the saddle point method one can obtain 
\begin{lem}\label{D(z)-asymp}
$D(z;\Gamma)$ admits the following asymptotic
\bea
D(z;\Gamma)\sim d(z)=1+\sum_{k=1}^\infty \frac{d_k}{z^{(r+1)k}}, \quad z\rightarrow \infty, \, z\in \omega^{-1} S_\Gamma.
\eea
\end{lem}

\section{A particular solution to the extended $r$-reduced KP hierarchy}
In this section we finally  prove Theorem \eqref{th1.1}. Let us further introduce some useful notations and lemmas.
\paragraph{Notations} 
Denote the vector space of formal series with finitely many  terms with positive powers by:
\bea 
\mathcal{A}=\left\{\sum_{n\geq  -m}  \frac{b_n}{z^{n}}\, \bigg| \,m\in\mathbb{Z},\,b_n \in\mathbb{C}\right\}.
\eea
For any $r\geq 2,$ let $\theta=e^{\frac{2\pi i}{r+1}}$ and let
\bea 
\mathcal{A}^{(k)}:=\left\{b(z)\in\mathcal{A}\, \big| \,b(\theta z)=\theta^k \, b(z)\right\},~k\in\mathbb{Z}. 
\eea
We also use subscripts to denote the leading order term, for example
$$
\mathcal{A}^{(0)}_1=\left\{1+\sum_{n\geq 1} \frac{ b_n } { z^{n(r+1)} }\, \bigg| \, b_n \in\mathbb{C} \right\},~\mathcal{A}^{(1)}_{3z^m}=\left\{3z^m+\sum_{n\geq 1} \frac{ b_n } { z^{n(r+1)-m} }\, \bigg| \, b_n \in\mathbb{C} \right\},~etc.
$$
Clearly $\mathcal{A}^{(k)}=z^k \mathcal{A}^{(0)},~\mathcal{A}^{(k)}=\mathcal{A}^{(k+r+1)},\quad \forall\,k\in\mathbb{Z} .$

\begin{lem} \label{Sz-period}
For any $k\in \mathbb{Z},\,S_z \mathcal{A}^{(k)}=\mathcal{A}^{(k-r)}, ~S_z^{\star} \mathcal{A}^{(k)}=\mathcal{A}^{(k-r)}.$
\end{lem}
\begin{prf} It suffices to consider the action of $S_z$ on a monomial $z^k:$
\bea S_z z^k=\left(k-\frac{r-1}2\right)z^{k-r}-z^{k+1}=z^{k-r}\left(\frac{2k-r+1}{2}-z^{r+1}\right). \eea
Similarly for $S_z^\star$.
\end{prf}

In the last section, we have defined a formal series $a(z)$ as the asymptotic expansion of $A(z;\Gamma)$ in $S_\Gamma.$ 
\begin{lem} \label{a}
$a(z)$ is the unique formal solution in $\mathcal{A}_1$ to the equation 
\bea S_z^r\,a(z)=(-z)^r\,a(z). \eea
\end{lem}
\begin{prf}
The uniqueness follows from the recursive procedure given by the ODE.
\end{prf}

Lemma \eqref{Sz-period} and Lemma \eqref{a} immediately imply
\begin{lem} For any $n\geq 0,\,S_z^n a(z)\in \mathcal{A}^{(n)}_{(-1)^n z^n}.$
\end{lem}

\begin{rmk}
We know from Lemma \eqref{Taylor-ff*} and the above lemmas that the point in  Sato's (formal)  Grassmannian corresponding to $Z(t)$ is given by
\bea W={\rm Span}_{\mathbb C} \{a(z),S_z a(z),S_z^2a(z),...\}.\eea
This is actually well-known for example, in \cite{KS} and in \cite{AM}.
\end{rmk}

\begin{lem}\label{orthogonal}
The following identities hold true:
\bea \label{ortho}
\frac{1}{2\pi i}\oint_{z=\infty} {S_z^\star}^m\left(a(\omega z)\right)\,S_z^n a(z)\,dz=0,\quad \forall\, m, n\geq 0.
\eea
If $b(z)\in\mathcal{A}_1$ satisfies that for all $n\geq 0$
\bea \label{ortho-condition}
\frac{1}{2\pi i}\oint_{z=\infty} b(z)\,S_z^n a(z)\,dz=0,
\eea
then $b(z)=a(\omega z).$
\end{lem}
\begin{prf}
Due to Lemma \ref{Taylor-ff*}, the identity \eqref{ortho} is nothing but the bilinear identity \eqref{bilinear-psipsi'} evaluated at $t|_{KP,\geq 2}=t'|_{KP,\geq 2}=0$. Write 
\bea b(z)=1+\sum_{k=1}^\infty \frac{b_k}{z^k}; \eea
then $b_1,b_2,b_3,...$ can be solved recursively from the relations \eqref{ortho-condition} by taking $n=1,2,3,...$ respectively.
\end{prf}

Before proceeding further,  let us mention that Lemma \ref{concomitant} and Lemma \ref{a} imply
\begin{lem}  The following identities hold true in $\mathcal{A}$:
\bea
\sum_{k=0}^{r-1} {S_z^\star}^{r-1-k} \left(a(\omega z)\right) S_z^k a(z)=(-1)^{r-1}r \, z^{r-1}.
\eea
\end{lem}
\begin{rmk}
In the case $r=2$, the identity appearing in this lemma was observed in \cite{Pixton} and it was used by Buryak \cite{Buryak2} to prove the analog of Lemma \ref{orthogonal} for $r=2$. We point out here that, in fact, this identity is a simple consequence of a classical object, the bilinear concomitant of adjoint equations, introduced by Legendre \cite{Ince}!
\end{rmk}

\begin{lem}\label{d(z)} The formal series $d(z)$ defined in the previous section as the asymptotic expansion of $D(z;\Gamma)$ in $\omega^{-1}S_\Gamma$ is the unique formal solution in $\mathcal{A}_1$ to the equation
\bea
-S_z^\star \left(\frac{d(z)}{z}\right)=a(\omega z).
\eea
\end{lem}
\begin{prf} The uniqueness of the solution in $\mathcal{A}_1$ follows from recursive procedure given by the above ODE. 
\end{prf}

\begin{emp}
For $r=2$ we have
\bea
a(z)&=&1-\frac{5}{24 z^3}+\frac{385}{1152 z^6}-\frac{85085}{82944 z^9}+\frac{37182145}{7962624 z^{12}}+\mathcal{O}(z^{-15}),\\
d(z)&=&1+\frac{41}{24 z^3}+\frac{9241}{1152z^6}+\frac{5075225}{82944z^9}+\frac{5153008945}{7962624 z^{12}}+\mathcal{O}(z^{-15}).
\eea
For $r=3$ we have
\bea
a(z)&=&1-\frac{7}{12z^4}+\frac{385}{288z^8}-\frac{39655}{10368 z^{12}}+\frac{665665}{497664 z^{16}}+\mathcal{O}(z^{-20}),\\
d(z)&=&1+\frac{31}{12z^4}+\frac{4849}{288z^8}+\frac{1785295}{10368 z^{12}}+\frac{1200383905}{497664z^{16}}+\mathcal{O}(z^{-20}).
\eea
For $r=4$ we have
\bea
a(z)&=&1-\frac{9}{8z^5}+\frac{441}{128 z^{10}}-\frac{30303}{5120 z^{15}}-\frac{25162137}{163840z^{20}}+\mathcal{O}(z^{-25}),\\
d(z)&=&1+\frac{29}{8z^5}+\frac{3921}{128z^{10}}+\frac{1990803}{5120z^{15}}+\frac{1089687543}{163840z^{20}}+\mathcal{O}(z^{-25}).
\eea
\end{emp}

\begin{thm} \label{esop}
Let
\begin{equation}\label{ZE}
Z_E(t_{_{KP}} )=Z(t)\,\Psi(t_{_{KP}} ),
\end{equation} be the partition function 
of the extended $r$-reduced KP hierarchy, where
$Z(t)$ is the partition function of the $A_{r-1}$ Frobenius manifold and 
\begin{equation}\label{PSI-Z}
\Psi(t_{_{KP}} )=\frac{1}{2\pi i}\,\oint_{z=\infty} g(z)\, \psi(t_{_{KP}} ;z)\,\frac{dz}{z}
\end{equation}
with 
\bea \label{req-g}
g(z)=1+\sum_{k\geq 1} \frac{g_k}{z^k}.
\eea
Then there exists a unique sequence of numbers $g_1,g_2,g_3,...$ such that 
\bea L_{-1}^{ext}Z_E(t_{_{KP}} )=0. \eea
And these numbers are given by 
\bea g_{(r+1)k+p}=0~(k\geq 0,\,p=1,2,...,r);\quad g_{(r+1)k}=d_k,\,k\geq 1.\eea
\end{thm}

\begin{prf}
Let $K(t_{_{KP}} ; z)$ denote the wave potential associated to $Z(t).$ Then by definition
\bea
Z_E(t_{_{KP}} )=\frac{1}{2\pi i}\oint_{z=\infty} g(z) K(t_{_{KP}} ;z)\,\frac{dz}{z}.
\eea
By using Lemma \ref{string-action-wave} we have
\bea
\label{string-sz}
0&=&\,L_{-1}^{ext} Z_E=\frac{1}{2\pi i}\oint_{z=\infty} g(z) L_{-1}^{ext}K(t_{_{KP}} ;z)\,\frac{dz}{z}=\,\frac{1}{2\pi i}\oint_{z=\infty} g(z)\, S_z (K(t_{_{KP}} ;z))\,\frac{dz}{z}\nn\\
&=&\,\frac{Z(t)}{2\pi i}\oint_{z=\infty} g(z)\, \,S_z (\psi(t_{_{KP}} ;z))\,\frac{dz}{z}.
\eea
Taking $t_1=t_2=...=0$ in the above equation we obtain
\begin{equation}\label{string-sz-initial}
\frac{1}{2\pi i}\oint_{z=\infty} g(z)\, S_z f(x;z)\,\frac{dz}z=0.
\end{equation}
Substituting \eqref{Taylor-A} in \eqref{string-sz-initial} we have
\bea \sum_{n=0}^\infty \frac{(-1)^n}{n!} \frac{1}{2\pi i}\oint_{z=\infty} S_z^\star \left(\frac{g(z)}{z}\right)  S_z^n a(z)\,dz\,x^n=0.\eea
This is equivalent to
\bea \frac{1}{2\pi i}\oint_{z=\infty} S_z^\star \left(\frac{g(z)}{z}\right)  S_z^n a(z)\,dz=0,\quad \forall\, n\geq 0. \eea
Due to Lemma \ref{orthogonal} and the requirement \eqref{req-g}, we obtain 
\bea S_z^\star \left(\frac{g(z)}{z}\right)=-a(\omega z).\eea
Lemma \ref{d(z)} implies 
\bea g(z)=d(z).\eea
Now if $g(z)=d(z)$, then we know from Lemma \ref{orthogonal} that \eqref{string-sz-initial} holds true.  Since the wave function $\psi(t_{_{KP}} ;z)$ satisfies the linear evolutionary equations \eqref{ET}, we know that \eqref{string-sz-initial} also implies  \eqref{string-sz}. The theorem is proved.
\end{prf}

\begin{prp}\label{initial-PSI}
The function $\Psi(t_{_{KP}} )$ defined by \eqref{PSI-Z} with $g(z)=d(z)$ satisfies 
\bea \Psi(t_{_{KP}} )|_{t_{\geq 2}=0}\equiv 1,\ \ \ \forall \, t_1.\eea
\end{prp}
\begin{prf}
Reminding the reader that $t_1 =x$, we have for all $x$
\bea
\Psi(t_{_{KP}} )|_{t_{\geq 2}=0}&=&\frac{1}{2\pi i}\,\oint_{z=\infty} d(z)\, f(x; z)\, \frac{dz}z\\
&=&\frac{1}{2\pi i}\,\oint_{z=\infty} d(z)\, \sum_{n=0}^\infty \frac{(-1)^n}{n!} S_z^n a(z)\, x^n\, \frac{dz}z=1.
\eea
The last equality follows from Lemma \ref{d(z)} together with  Lemma \ref{orthogonal} for the terms $n\geq 1$,  while for $n=0$ we use the 
fact that both $d(z)$ and  $a(z)$ are of the form $1 + \mathcal O(z^{-r-1})$.
The proposition is proved.
\end{prf}

Theorem \eqref{th1.1} is then a combination of Theorem \eqref{esop} and Proposition \eqref{initial-PSI}.

\paragraph{Acknowledgements}
We thank Ferenc Balogh and Thomas Bothner for helpful discussions. M. B. acknowledges the support of the {\em Natural Sciences and Engineering Research Council of
Canada} and the {\em Fonds de recherche du  Qu\'ebec -Nature et technologies}. 
D. Y. wishes to thank Youjin Zhang and Boris Dubrovin for their advises and helpful discussions. He acknowledges the support of PRIN 2010-11 Grant ``Geometric and analytic theory of Hamiltonian systems in finite and infinite dimensions" of Italian Ministry of Universities and Researches, and the support of the Marie Curie IRSES project RIMMP.  He also wishes to thank the Centre de Recherches Math\'emetiques and the Department of Mathematics and Statistics at Concordia University for generous hospitality, where the manuscript was completed.

\end{document}